\documentclass[11 pt]{article} 

\usepackage{amssymb, amsmath, amsthm, bbm}
\usepackage{fullpage} 
\usepackage{hyperref}

\theoremstyle{definition}
\newtheorem{defn}{Definition}

\theoremstyle{plain}
\newtheorem{thm}[defn]{Theorem}

\newtheorem{cor}[defn]{Corollary}

\begin{document}

\title{A Matricial Algorithm for Polynomial Refinement}
\author{Emily J. King}

\maketitle

\abstract{In order to have a multiresolution analysis, the scaling function must be refinable.  That is, it must be the linear combination of 2-dilation, $\mathbb{Z}$-translates of itself.  Refinable functions used in connection with wavelets are typically compactly supported.  In 2002, David Larson posed the question in his REU site, ``Are all polynomials (of a single variable) finitely refinable?"  That summer the author proved that the answer indeed was true using basic linear algebra.  The result was presented in a number of talks but had not been typed up until now.  The purpose of this short note is to record that particular proof.}

\section*{Polynomial Refinement}
A scaling function for a multiresolution analysis must be be \emph{$2$-refinable}, but not all refinable functions can be a scaling function.
\begin{defn}
A function $f: \mathbb{R} \rightarrow \mathbb{C}$ is \emph{($2$-)refinable} if there exists a sequence $\alpha \in \ell^\infty(\mathbb{Z})$ such that 
\begin{equation}\label{eqn:ref}
f(\cdot) = \sum_{\ell \in \mathbb{Z}} \alpha_\ell f(2 \cdot - \ell).
\end{equation}
If $\alpha$ is finitely supported, we say that $f$ is \emph{finitely ($2$-)refinable}.  When $2$ is replaced by $a \neq 0$ in (\ref{eqn:ref}), we say that $f$ is \emph{$a$-refinable}.
\end{defn}
The focus on $a = 2$ is due to the early connection of refinability with wavelet theory.  In approximation theory and signal processing the $\alpha$ in (\ref{eqn:ref}) is called the \emph{mask} or \emph{low pass filter sequence}, respectively.
\begin{thm}
Let $p:\mathbb{R} \rightarrow \mathbb{C}$ be a polynomial of degree $n$.  Let $\{ \ell_i \}_{i = 0}^{n}$ be any set of $n+1$ distinct integers.  For any $a \neq 0$, there exists $\alpha \in \ell^\infty(\mathbb{Z})$ supported in $\{ \ell_i \}$ such that
\begin{equation}\label{eqn2}
p(\cdot) = \sum_{i = 0}^{n} \alpha_{\ell_i} p(a \cdot -\ell_i).
\end{equation}
\end{thm}

The proof is at its core the same as the proof of Theorem 1.8 in \cite{undergrads}, which was independently discovered by Gustafson, Savir, and Spears a few years after the author of this note.  However, they present it in a slightly different light in their paper and focus on the end result of refinability rather than the fact that for any polynomial any sequence of $n+1$ distinct shifts is associated with a refinement mask.  Note that we will index our matrices and vectors starting with $0$ to make the notation easier.

\begin{proof}
Since the polynomial is of degree $n$, it may be written in the form $p(x) = \sum_{k = 0}^n p_k x^k$ with $p_n \neq 0$.  Plugging this into (\ref{eqn2}), we obtain 
\begin{eqnarray*}
p(x) & = & \sum_{i = 0}^n \alpha_{\ell_i} \sum_{k=0}^n p_k (ax - \ell_i)^k \\
& = & \sum_{i = 0}^n \alpha_{\ell_i} \sum_{k=0}^n p_k \sum_{j=0}^k \binom{k}{j}a^j x^j (-\ell_i)^{k-j} \\
& = & \sum_{j=0}^n  x^j a^j \sum_{k=j}^n p_k  \binom{k}{j} \sum_{i = 0}^n \alpha_{\ell_i}  (-\ell_i)^{k-j}. 
\end{eqnarray*} 
A comparison of the monomial coefficients followed by a substitution ($\overline{k} = n-k+j$) yields
\begin{eqnarray*}
p_j & = & a^j \sum_{k=j}^n p_k  \binom{k}{j} \sum_{i = 0}^n \alpha_{\ell_i}  (-\ell_i)^{k-j} \quad \textrm{for each $0 \leq j \leq n$}\\
& = & a^j \sum_{k=j}^n p_{n-k+j}  \binom{n-k+j}{j} \sum_{i = 0}^n \alpha_{\ell_i}  (-\ell_i)^{n-k} \quad \textrm{for each $0 \leq j \leq n$}.
\end{eqnarray*}
This may be rewritten as a matrix equation
\begin{equation}\label{eqn:matrix}
p = D_a C V m,
\end{equation}
where $D_a$ is the diagonal matrix with $(D_a)_{i,i} = a^{i}$ for $0 \leq i \leq n$, $C$ is the invertible upper triangular matrix
\begin{equation*}
C = \left( \left\{ \begin{array}{ccc}p_{n-j+i}\binom{n-j+i}{j} & ; & i \leq j \\ 0 & ; & \textrm{else} \end{array}\right. \right)_{0 \leq i,j \leq n},
\end{equation*}
$V$ is the invertible (since the shifts $\ell_i$ are distinct) Vandermonde matrix
\begin{equation*}
V = \left((-\ell_j)^{n-i} \right)_{0 \leq i,j \leq n},
\end{equation*}
and $m$ is the refinement mask $m = ( \alpha_{\ell_i})_{i = 0}^n$.
\end{proof}

\begin{cor}
All polynomials $p:\mathbb{R} \rightarrow \mathbb{C}$ are finitely $2$-refinable.
\end{cor}
Henning Thielemann also independently proved this fact using different methods in \cite{Henning}.

Much is known about Vandermonde matrices, in fact they admit a well-known $LU$-decomposition \cite{Vander}.  Thus (\ref{eqn:matrix}) also provides a quick algorithm for determining refinement masks of polynomials.


\begin{thebibliography}{c}
\bibitem[1]{undergrads} Gustafson, Paul; Savir, Nathan; Spears, Ely (2006-11-14), ``A Characterization of Refinable Rational Functions," \emph{American Journal of Undergraduate Research}, \textbf{5}(3): 11Ð20
\bibitem[2]{Henning} Thielemann, Henning (2011-04-13). "Polynomial Functions are Refinable,"  accepted \emph{International Journal of Wavelets, Multiresolution and Information Processing}.
\bibitem[3]{Vander} Turner, L. Richard (1966), ``Inverse of the Vandermonde Matrix with Applications,'' \emph{NASA Technical Note}.
\end{thebibliography}
\end{document}